\documentclass{aa}

\usepackage{graphicx}
\usepackage{txfonts}

\newcommand{\Msol}{$M_{\odot}$}
\newcommand{\so}{$\sigma$~Orionis}

\begin{document}

   \title{A search for substellar members in the Praesepe and \so~clusters }
   \titlerunning{A search for substellar members in the Praesepe and \so~clusters}

   \subtitle{}

   \author{B. M. Gonz\'alez-Garc\'{\i}a \inst{1,2}
	  \and M. R. Zapatero Osorio \inst{1,3}
          \and V. J. S. B\'ejar \inst{3}
          \and G. Bihain \inst{3,4}
          \and D. Barrado y Navascu\'es \inst{1}
          \and J. A. Caballero \inst{3}
          \and M. Morales-Calder\'on \inst{1}
          }

   \offprints{Gonz\'alez-Garc\'{\i}a, Zapatero Osorio, bmgg@laeff.inta.es,mosorio@iac.es}
  
   \institute{Laboratorio de Astrof\'\i sica Espacial y F\'\i sica Fundamental (LAEFF-INTA), P.\,O$.$ 50727, E-28080 Madrid, Spain
	      \and
              XMM-Newton Instrument Controller at ESAC, INSA, Madrid, Spain
              \and
              Instituto de Astrof\'\i sica de Canarias (IAC), V\'\i a L\'actea s/n, E-38205 La Laguna, Tenerife, Spain
              \and
              Consejo Superior de Investigaciones Cient\'\i ficas (CSIC), Spain
   }

   \date{Received ; accepted }

   \abstract{ We have conducted deep photometric searches for
     substellar members of the Praesepe (0.5--1~Gyr) and \so~(3~Myr)
     star clusters using the Sloan $i'$ and $z'$ broad-band filters,
     with the 3.5-m and the 5-m Hale telescopes on the Calar Alto and
     Palomar Observatories, respectively. The total area surveyed was
     1177 arcmin$^2$ and 1122 arcmin$^2$ towards the central regions
     of Praesepe and \so, respectively. The 5-$\sigma$ detection limit
     of our survey is measured at $i'$\,=\,24.5 and $z'$\,=\,24 mag,
     which according to state-of-the-art evolutionary models
     corresponds to masses of 50--55~$M_{\rm Jup}$ (Praesepe) and
     6~$M_{\rm Jup}$ (\so), i.e., well within the substellar
     regime. Besides recovering previously known cluster members
     reported in the literature, we have identified new photometric
     candidates in both clusters whose masses expand the full range
     covered by our study. In \so, follow-up near-infrared photometry
     has allowed us to confirm the likely cluster membership of three
     newly discovered planetary-mass objects. The substellar mass
     function of \so, which is complete from the star--brown dwarf
     borderline down to 7~$M_{\rm Jup}$, keeps rising smoothly with a
     slope of $\alpha$\,=\,0.6\,$^{+0.5}_{-0.1}$ (d$N$/d$M$ $\sim$
     $M^{-\alpha}$). Very interestingly, one of the faintest Praesepe
     candidates for which we have also obtained follow-up
     near-infrared $JHK_s$ photometry nicely fits the expected optical
     and infrared photometric sequence of the cluster. From its
     colors, we have estimated its spectral type to be between L4 and
     L6. If confirmed as a true Praesepe member, it would become the
     first L-type brown dwarf (50--60~$M_{\rm Jup}$) identified in an
     intermediate-age star cluster. Our derivation of the Praesepe
     mass function, which is based on state-of-the-art evolutionary
     models, depends strongly on the cluster age. For the youngest
     possible ages (500--700~Myr), our results suggest that there is a
     deficit of Praesepe brown dwarfs in the central regions of the
     cluster, while the similarity between the Praesepe and \so~mass
     functions increases qualitatively for models older than 800~Myr.

   }

\maketitle

\begin{figure*}   
\centering
\caption{The Palomar (rectangles) and Calar (squares) photometric
  surveys carried out in the Praesepe (left panel) and \so~(right
  panel) star open clusters. The location of our candidates is
  indicated with symbols: solid squares correspond to candidates with
  high photometric cluster membership probability, and open circles
  stand for low probability candidates. Bottom images are taken from
  the Digital Sky Survey (DSS1, red data).  [Provided in jpeg format in astro-ph.]}
\label{survey_photo}
\end{figure*}

\section{Introduction}

The knowledge of the substellar mass function of various star-forming regions and very young open clusters ($\le$\,10~Myr) can provide key constrains on our understanding of the formation mechanisms giving rise to very low-mass stars and brown dwarfs (masses below 75~$M_{\rm jup}$, or 0.072~\Msol, Chabrier \& Baraffe \cite{chabrier00}). The study of intermediate-age star clusters (100--1000~Myr) is relevant to track down the dynamical evolution of brown dwarfs and its impact on the shape of the mass function. It is also important to detail the brown dwarfs luminosity and effective temperature changes with time. Furthermore, ultracool L- and T-type brown dwarf members of intermediate-age star clusters would become benchmark objects to understand the very low-mass stellar and substellar populations in the solar neighborhood. 

In this paper, we present the results of deep photometric surveys in two open clusters of different ages: \so~and Praesepe. Our main objectives are: the identification, for the first time, of the Praesepe unambiguous substellar population, and the extension of our previous deep explorations carried out in \so~to search for planetary-mass candidates below the deuterium burning-mass limit (Zapatero Osorio et al. \cite{osorio00}). Both clusters have roughly solar metallicity (Caballero \cite{caballero06}; Hambly et al. \cite{hambly95a}) and are characterized by very little internal extinction (Lee \cite{lee68}; Crawford \& Barnes \cite{crawford69}). The Praesepe cluster has an age in the interval 0.5--1~Gyr, and is located at a distance of 180~pc (Robichon et al. \cite{robichon99}), while the youngest cluster, \so~(1--8~Myr, Oliveira et al. \cite{oliveira02}; Zapatero Osorio et al. \cite{osorio02}, with a likely mean age around 3~Myr) lies at 352~pc (Perryman et al. \cite{perryman97}). 

Previous large scale proper motion studies of Praesepe (Hambly et al. \cite{hambly95b}; Adams et al. \cite{adams02}) covered most of the lower main sequence of the cluster down to 0.1~$M_{\odot}$. Deeper surveys by Pinfield et al. (\cite{pinfield97}, \cite{pinfield00}), Magazz\`u et al. (\cite{magazzu98}), and Chappelle et al. (\cite{chappelle05}) have extended the Praesepe population down to spectral types M9--L0, which corresponds to a mass of 0.06--0.08~$M_{\odot}$, i.e., rather close to the substellar borderline. Chappelle et al. (\cite{chappelle05}) concluded that Praesepe substellar members are likely depleted from the cluster due to dynamical evolution. Our survey does not cover a significant area of Praesepe (thought to be of several tens square degrees), but to the best of our knowledge, is the deepest photometric exploration carried out in the cluster central regions so far. 

We present our optical surveys and cluster member candidate selection criteria in Section~2. Follow-up near-infrared photometry (also presented in Section~2) allows us to refine photometric membership status of the candidates. In Section~3 we discuss on the mass of the selected candidates, derive the Praesepe and \so~mass functions, and compare them to various mass functions available in the literature. We find new planetary-mass candidates (6--13~$M_{\rm Jup}$) in \so, and one likely Praesepe brown dwarf (50--60~$M_{\rm Jup}$) with colors typical of L4--L6 spectral types.


\section{Observations}

\subsection{Optical photometry}

We have obtained deep CCD images of the young star clusters Praesepe and \so~using the Sloan $i'$ and $z'$ filters and the 5-m Hale Telescope on the Palomar Observatory (2002 February 07) and the 3.5-m telescope on the Calar Alto Observatory (2003 December 23, 24). Both telescopes were equipped with large format multi-CCD cameras (the Large Field Camera, LFC, at Palomar and the Large Area Imager, LAICA, at Calar Alto), which cover a substantial area on the sky in one single shot. Instrumental details like pixel size, number of CCDs, total area covered by each instrument, filter central wavelengths and passbands are summarized in Table~\ref{instruments}. The two filter sets show similar properties. All observations were carried out under clear transparency conditions and with seeing values of 1\farcs6--1\farcs7\ (Palomar) and 1\farcs7--1\farcs8 (Calar).

\begin{table*}
\caption{Optical instruments employed in our survey.}
\label{instruments}
\centering           
\begin{tabular}{lccccccccc}     
\hline    
\hline    
\multicolumn{6}{c}{} &
\multicolumn{2}{c}{Sloan $i'$ filter} &
\multicolumn{2}{c}{Sloan $z'$ filter} \\
Telescope  & Observatory& Camera             & Detectors                   & Pixel size   & Area        & $\lambda$$_{\rm c}$ & $\Delta \lambda$ & $\lambda$$_{\rm c}$ & $\Delta \lambda$ \\
	   &            &                    &                             &(\arcsec/pix) &(arcmin$^2$) & (nm)                & (nm)             & (nm)                & (nm)             \\ 
\hline      
5-m Hale   & Palomar    & LFC$^{\mathrm{a}}$ & 6$\times$(1024$\times$2048) & 0.360        & 450         & 768                 & 154              & 900                 & 180              \\
3.5-m CAHA & Calar Alto & LAICA              & 4$\times$(4096$\times$4096) & 0.225        & 940         & 770                 & 150              & 920                 & 160              \\
\hline
\end{tabular}
\begin{list}{}{}
\item[$^{\mathrm{a}}$] The original size of each CCD detector is 2048$\times$4096, and the pixel size is 0\farcs18. We applied a 2$\times$2 binning to all LFC CCDs.
\end{list}
\end{table*}

\begin{table*}
\caption{Log of optical observations.}
\label{optlog}
\centering           
\begin{tabular}{lllcrcc}
\hline    
\hline    
Cluster & Survey & Obs. date     & Filter & \multicolumn{1}{c}{Exp.} & $m_{\rm comp}$, $m_{\rm lim}$ & Airmass \\
        &        & (UT)          &        & \multicolumn{1}{c}{(s)}  & (mag)                         &         \\
                         \hline    
\so     & Palomar& 2002 Feb 07   & $i'$   & 46$\times$200            & 23.2, 24.5                    & 1.38--1.39 \\ 
        & Palomar& 2002 Feb 07   & $z'$   & 21$\times$200            & 22.5, 24.0                    & 1.40--1.93 \\ 
        & Calar  & 2003 Dec 23   & $i'$   & 26$\times$500            & 23.8, 24.5                    & 1.30--1.93 \\ 
        & Calar  & 2003 Dec 24   & $z'$   & 26$\times$600            & 22.8, 24.0                    & 1.30--1.95 \\ 
Praesepe& Palomar& 2002 Feb 07   & $i'$   & 28$\times$200            & 23.8, 24.8                    & 1.03--1.23 \\ 
        & Palomar& 2002 Feb 07   & $z'$   & 21$\times$200            & 23.3, 24.3                    & 1.25--1.80 \\ 
        & Calar  & 2003 Dec 23   & $i'$   &  9$\times$500            & 23.3, 24.5                    & 1.05--1.13 \\ 
        & Calar  & 2003 Dec 23-24& $z'$   & 18$\times$500            & 22.8, 24.0                    & 1.05--1.44 \\ 
\hline    
\end{tabular}
\end{table*}

The log of the optical observations is shown in Table~\ref{optlog}, where we provide the total integration time per filter (i.e., number of frames multiplied by individual exposure time) and the airmass intervals in which data were collected. Data were acquired by applying small telescope offsets (dithers of 9\arcsec~to 12\arcsec) from one exposure to the next. This allowed us to create a nightly super-flat-field image as the result of the median of all the science frames taken in the same filter. Raw data were bias substracted and divided by the corresponding normalized superflats. Strong fringing patterns were apparent in the $z'$-band filter; they were corrected down to a level of 0.5\%~with the superflats. We aligned all individual flat-fielded frames and stacked them altogether to produce the final deep $i'$- and $z'$-band images free of cosmic rays. Point-like sources were identified using the Sextractor software (clearly resolved objects were mostly avoided). Some faint sources close to the detection limit that were missed by Sextractor were added manually after visual inspection of the images.

We depict in Fig.~\ref{survey_photo} the regions explored in the Praesepe and \so~clusters. There is an overlap of 70 arcmin$^2$ (Praesepe) and 125 arcmin$^2$ (\so) between the Palomar and Calar surveys. By taking into account these overlapping regions and the dither pattern of the observations, the final area covered by our optical survey down to the faintest detection limit is 1177 arcmin$^2$ in the Praesepe cluster and 1122 arcmin$^2$ in \so, which represents about 2\%~and 30\%~of the total cluster areas, respectively. The Calar survey imaged regions around the accepted centers of mass of the two clusters, while the Praesepe Palomar survey was centered close to the location of the previously known member Roque Praesepe~1 (RPr\,1, Magazz\`u et al. \cite{magazzu98}), and the Palomar search mapped the Southeast area of \so, in good agreement with the survey of Caballero (\cite{tesis06}) and Caballero et al. (\cite{caballero06}).

\begin{table*}
\caption{Previously known Praesepe candidates in our survey.}
\label{knownprae}           
\centering           
\begin{tabular}{l c c c l}  
\hline    
\hline        
Object    & $z'$           & $i'-z'$       & $I^{\mathrm{a}}$ & Reference \\
\hline
WFC\,210  & 17.58$\pm$0.20 & 1.13$\pm$0.28 & 18.02 & Chappelle et al $.$ (\cite{chappelle05}) \\
PHJ\,20   & 17.65$\pm$0.20 & 1.21$\pm$0.28 & 18.48 & Hodgkin et al$.$ (\cite{hodgkin99}) \\ 
WFC\,193  & 17.81$\pm$0.20 & 1.13$\pm$0.28 & 18.21 & Chappelle et al $.$ (\cite{chappelle05}) \\
PHJ\,18   & 18.43$\pm$0.20 & 1.64$\pm$0.28 & 19.63 & Hodgkin et al$.$ (\cite{hodgkin99}) \\ 
WFC\,88   & 19.06$\pm$0.20 & 1.51$\pm$0.29 & 19.83 & Chappelle et al $.$ (\cite{chappelle05}) \\
RPr\,1    & 19.71$\pm$0.20 & 1.53$\pm$0.29 & 21.01 & Magazz\`u et al $.$ (\cite{magazzu98}) \\
WFC\,60   & 19.86$\pm$0.20 & 1.66$\pm$0.28 & 20.79 & Chappelle et al $.$ (\cite{chappelle05}) \\
WFC\,24   & 20.32$\pm$0.20 & 1.66$\pm$0.29 & 21.19 & Chappelle et al $.$ (\cite{chappelle05}) \\
\hline
\end{tabular}
\begin{list}{}{}
\item[$^{\mathrm{a}}$] $I$-band magnitude from the literature.
\end{list}
\end{table*}

\begin{table*}
\caption{Previously known \so~candidates in our survey.}
\label{knownso}           
\centering           
\begin{tabular}{l c c c l}     
\hline    
\hline    
Object                                   & $z'$            & $i'-z'$       & $I^{\mathrm{a}}$ & Reference\\
\hline
S\,Ori\,13                               & 15.66$\pm$0.21  & 1.05$\pm$0.30 & 16.05 & B\'ejar et al$.$ (\cite{bejar06})\\
S\,Ori\,14                               & 15.92$\pm$0.21  & 1.26$\pm$0.29 & 16.35 & B\'ejar et al$.$ (\cite{bejar99})\\
S\,Ori\,19                               & 16.11$\pm$0.21  & 1.08$\pm$0.29 & 16.62 & B\'ejar et al$.$ (\cite{bejar06})\\
S\,Ori\,23                               & 16.43$\pm$0.21  & 1.14$\pm$0.29 & 16.88 & B\'ejar et al$.$ (\cite{bejar06})\\
S\,Ori\,24                               & 16.48$\pm$0.20  & 1.10$\pm$0.29 & 16.90 & B\'ejar et al$.$ (\cite{bejar06})\\
S\,Ori\,28                               & 16.74$\pm$0.21  & 1.24$\pm$0.29 & 17.23 & B\'ejar et al$.$ (\cite{bejar06})\\
S\,Ori\,J053922.2$-$024552               & 16.78$\pm$0.21  & 1.08$\pm$0.29 & 17.05 & Caballero et al$.$ (\cite{caballero04})\\
S\,Ori\,32	                         & 16.89$\pm$0.21  & 1.26$\pm$0.29 & 17.40 & B\'ejar et al$.$ (\cite{bejar06})\\
KNJ2005 72                               & 17.01$\pm$0.21  & 1.31$\pm$0.29 & 17.82 & Kenyon et al $.$ (\cite{kenyon05})\\
KNJ2005 70                               & 17.13$\pm$0.21  & 1.29$\pm$0.29 & 17.81 & Kenyon et al $.$ (\cite{kenyon05})\\
KNJ2005 71                               & 17.26$\pm$0.21  & 1.12$\pm$0.29 & 17.82 & Kenyon et al $.$ (\cite{kenyon05})\\
S\,Ori\,38	                         & 17.34$\pm$0.24  & 1.32$\pm$0.35 & 17.70 & B\'ejar et al$.$ (\cite{bejar06})\\
S\,Ori\,J053926.8$-$022614               & 17.69$\pm$0.21  & 1.15$\pm$0.30 & 18.25 & B\'ejar et al$.$ (\cite{bejar06})\\
S\,Ori\,J053918.1$-$025257$^{\mathrm{b}}$& 17.99$\pm$0.21  & 1.66$\pm$0.29 & 18.64 & Caballero et al$.$ (\cite{caballero04})\\
S\,Ori\,J053948.1$-$022914               & 18.06$\pm$0.21  & 1.20$\pm$0.30 & 18.52 & B\'ejar et al$.$ (\cite{bejar06})\\
S\,Ori\,42	                         & 18.33$\pm$0.21  & 1.70$\pm$0.30 & 19.04 & Caballero et al$.$ (\cite{caballero04})\\
S\,Ori\,44	                         & 18.94$\pm$0.21  & 1.29$\pm$0.29 & 19.39 & B\'ejar et al $.$ (\cite{bejar99})\\
S\,Ori\,45	                         & 18.99$\pm$0.21  & 1.58$\pm$0.30 & 19.49 & Caballero et al$.$ (\cite{caballero04})\\
S\,Ori\,J053929.4$-$024636               & 19.12$\pm$0.21  & 1.68$\pm$0.29 & 19.73 & Caballero et al$.$ (\cite{caballero04})\\
S\,Ori\,J053946.5$-$022423               & 19.09$\pm$0.21  & 1.59$\pm$0.29 & 20.14 & B\'ejar et al$.$ (\cite{bejar01})\\
S\,Ori\,52	                         & 19.76$\pm$0.21  & 1.78$\pm$0.29 & 20.55 & B\'ejar et al$.$ (\cite{bejar06})\\
S\,Ori\,53	                         & 20.42$\pm$0.22  & 1.85$\pm$0.31 & 20.96 & Caballero et al$.$ (\cite{caballero04})\\
S\,Ori\,J053956.8$-$025315               & 20.49$\pm$0.21  & 1.86$\pm$0.30 & 21.25 & Caballero et al$.$ (\cite{caballero04})\\
S\,Ori\,J053858.6$-$025228               & 21.28$\pm$0.24  & 1.98$\pm$0.35 & 22.17 & Caballero et al$.$ (\cite{caballero04})\\
S\,Ori\,67         	                 & 22.05$\pm$0.22  & 1.78$\pm$0.35 & 23.04 & B\'ejar et al$.$ (\cite{bejar06})\\
\hline
\end{tabular}
\begin{list}{}{}
\item[$^{\mathrm{a}}$] $I$-band magnitude from the literature.
\item[$^{\mathrm{b}}$] It is detected at both Palomar and Calar surveys. We provide the mean $i'$ and $z'$ photometry. The $i'$ magnitudes differ by 0.3~mag suggesting photometric variability and supporting the conclusions on the source variable nature reported by Caballero et al$.$ (\cite{caballero04}).
\end{list}
\end{table*}

Recently, Holland et al. (\cite{holland00}) have claimed that the present Praesepe cluster is the result of a recent merging of clusters with different ages. The ``subcluster'' mentioned by these authors lies about 3~pc ($\sim$1~deg) Southwest from the cluster centre. We note that our Praesepe survey does not overlap with it. Regarding \so, Jeffries et al. (\cite{jeffries06}) have found that there are two distinct kinematic groups between the stars $\sigma$ and $\zeta$ of Orion. Both groups are very young ($\le$10\,Myr) and located at a distance of 350--450~pc; the most numerous group is concentrated around \so, while the population of the second component increases towards the North. According to Fig.~2 of Jeffries et al. (\cite{jeffries06}), only the Northwest CCD of the Calar survey is located in a region where the number of young objects likely belonging to the second group is comparable to that of \so~true members. Thus, we do not expect a significant contamination (less than 10\%) in our \so~survey.

\begin{table*}
\caption{New Praesepe candidates from our survey}
\label{newprae}           
\centering           
\begin{tabular}{l c c c c c c c l}     
\hline    
\hline    
\multicolumn{3}{c}{} &
\multicolumn{2}{c}{Palomar survey} &
\multicolumn{2}{c}{Calar survey} &
\multicolumn{2}{c}{} \\
IAU name                 &  RA (J2000)& DEC (J2000) &  $z'$          &  $i'-z'$      &  $z'$         &  $i'-z'$     & $I^{\mathrm{a}}$ &Phot. Prob.\\
          & ($^h$ ~ $^m$ ~ $^s$)&($^o$ ~ $'$ ~ $''$)&                &               &               &              &     &           \\
\hline
Prae\,J083850.6$+$192317 & 08 38 50.55 & $+$19 23 16.5 &                &               & 17.57$\pm$0.20&1.08$\pm$0.28 &18.20&high\\
Prae\,J084051.2$+$191010 & 08 40 51.24 & $+$19 10 09.5 & 17.75$\pm$0.20 & 0.93$\pm$0.28 &               &              &18.41&low\\
Prae\,J084108.0$+$191401 & 08 41 08.00 & $+$19 14 00.6 & 18.42$\pm$0.20 & 1.04$\pm$0.28 &               &              &19.20&low\\
Prae\,J083905.4$+$192032 & 08 39 05.38 & $+$19 20 31.5 &                &               & 19.16$\pm$0.20&1.26$\pm$0.28 &20.06&low\\
Prae\,J084130.4$+$190449 & 08 41 30.43 & $+$19 04 48.5 & 19.21$\pm$0.20 & 1.16$\pm$0.28 &               &              &20.12&low\\
Prae\,J084115.8$+$191743 & 08 41 15.79 & $+$19 17 43.0 &                &               & 19.39$\pm$0.20&1.19$\pm$0.28 &20.33&low\\
Prae\,J084114.5$+$191110 & 08 41 14.46 & $+$19 11 10.1 & 19.42$\pm$0.20 & 1.14$\pm$0.28 &               &              &20.36&low\\
Prae\,J084126.5$+$195200 & 08 41 26.54 & $+$19 51 59.7 &                &               & 19.78$\pm$0.20&1.38$\pm$0.28 &20.78&low\\
Prae\,J084108.1$+$191111 & 08 41 08.13 & $+$19 11 10.8 & 20.07$\pm$0.20 & 1.42$\pm$0.29 &               &              &21.12&low\\
Prae\,J084039.3$+$192840 & 08 40 39.28 & $+$19 28 39.6 &                &               & 20.26$\pm$0.20&1.78$\pm$0.28 &21.34&high\\
Prae\,J084052.5$+$190504 & 08 40 52.52 & $+$19 05 03.5 & 20.73$\pm$0.20 & 1.42$\pm$0.29 &               &              &21.89&low\\
Prae\,J084122.6$+$192419 & 08 41 22.58 & $+$19 24 19.0 & 21.33$\pm$0.25 & 1.55$\pm$0.35 & 21.21$\pm$0.20&1.61$\pm$0.28 &22.52&low\\
Prae\,J084049.1$+$195913 & 08 40 49.14 & $+$19 59 12.6 &                &               & 21.30$\pm$0.20&1.71$\pm$0.29 &22.55&low\\
Prae\,J084044.9$+$200025 & 08 40 44.90 & $+$20 00 25.3 &                &               & 21.58$\pm$0.20&1.64$\pm$0.29 &22.88&low\\
Prae\,J084045.2$+$192532 & 08 40 45.24 & $+$19 25 32.3 &                &               & 21.72$\pm$0.20&1.68$\pm$0.29 &23.05&low\\
Prae\,J083910.2$+$195610 & 08 39 10.20 & $+$19 56 10.1 &                &               & 21.86$\pm$0.20&1.65$\pm$0.29 &23.21&low\\
Prae\,J084111.9$+$192043 & 08 41 11.90 & $+$19 20 43.3 & 22.33$\pm$0.22 & 1.79$\pm$0.32 & 22.33$\pm$0.20&2.27$\pm$0.43 &23.76&high\\
Prae\,J083904.7$+$194959 & 08 39 04.72 & $+$19 49 58.6 &                &               & 22.36$\pm$0.20&1.87$\pm$0.31 &23.80&low\\
Prae\,J084057.3$+$190258 & 08 40 57.37 & $+$19 02 57.7 & 22.49$\pm$0.27 & 1.84$\pm$0.35 &               &              &23.95&low\\
Prae\,J084111.2$+$191349 & 08 41 11.15 & $+$19 13 49.1 & 22.75$\pm$0.25 & 1.82$\pm$0.36 &               &              &24.25&low\\
\hline
\end{tabular}
\begin{list}{}{}
\item[$^{\mathrm{a}}$] Estimated from the $z'$ photometry.
\end{list}
\end{table*}

\begin{table*}
\caption{New \so~candidates.}
\label{newso}           
\centering           
\begin{tabular}{l c c c c c c c l}   
\hline    
\hline    
\multicolumn{3}{c}{} &
\multicolumn{2}{c}{Palomar survey} &
\multicolumn{2}{c}{Calar survey} &
\multicolumn{2}{c}{} \\
IAU name                 &RA (J2000)&DEC (J2000)&$z'$          &$i'-z'$      &$z'$          &$i'-z'$      & $I^{\mathrm{a}}$ &Phot. Prob.\\
      & ($^h$ ~ $^m$ ~ $^s$)&($^o$ ~ $'$ ~ $''$)&              &             &              &             &     &           \\
\hline
S\,Ori\,J053724.5$-$021856$^{\mathrm{b}}$ &05 37 24.46&$-$02 18 55.7&              &             &16.58$\pm$0.21&1.12$\pm$0.29&17.07&high      \\
S\,Ori\,J053855.0$-$024034 &05 38 54.95&$-$02 40 33.6&17.69$\pm$0.21&1.64$\pm$0.29 &              &             &18.24&high      \\
S\,Ori\,J053734.5$-$025559 &05 37 34.51&$-$02 55 58.5&              &              &17.79$\pm$0.21&1.35$\pm$0.30&18.35&high      \\
S\,Ori\,J053908.4$-$023902 &05 39 08.38&$-$02 39 01.9&18.03$\pm$0.22&1.22$\pm$0.31 &              &             &18.60&low       \\
S\,Ori\,J053912.6$-$025724 &05 39 12.60&$-$02 57 23.9&18.69$\pm$0.21&1.30$\pm$0.29 &              &             &19.29&low       \\
S\,Ori\,J053731.9$-$021634$^{\mathrm{b}}$ &05 37 31.89&$-$02 16 33.8&              &             &19.03$\pm$0.21&1.49$\pm$0.29&19.65&high      \\
S\,Ori\,J053946.3$-$022631 &05 39 46.33&$-$02 26 31.3&              &              &19.70$\pm$0.29&1.51$\pm$0.39&20.36&low       \\
S\,Ori\,J053913.9$-$021621$^{\mathrm{b}}$ &05 39 13.94&$-$02 16 21.2&              &             &19.80$\pm$0.21&1.73$\pm$0.29&20.47&high      \\
S\,Ori\,J053813.5$-$021934$^{\mathrm{b}}$ &05 38 13.54&$-$02 19 34.2&              &             &20.13$\pm$0.21&1.48$\pm$0.29&20.83&low       \\
S\,Ori\,J053758.0$-$025146 &05 37 57.98&$-$02 51 46.1&              &              &20.63$\pm$0.26&1.59$\pm$0.38&21.38&low       \\
S\,Ori\,J053957.4$-$025006 &05 39 57.40&$-$02 50 06.2&              &              &21.26$\pm$0.22&1.95$\pm$0.32&22.10&high      \\
S\,Ori\,J053919.2$-$025910 &05 39 19.23&$-$02 59 10.4&21.60$\pm$0.28&1.84$\pm$0.38 &              &             &22.50&low       \\
S\,Ori\,J054007.8$-$022234 &05 40 07.75&$-$02 22 34.3&              &              &21.65$\pm$0.22&1.79$\pm$0.33&22.56&low       \\
S\,Ori\,J053844.5$-$025514$^{\mathrm{c}}$ &05 38 44.54&$-$02 55 14.3&21.66$\pm$0.27&2.02$\pm$0.38&      &      &22.57&high      \\ 
S\,Ori\,J054008.5$-$024551$^{\mathrm{c}}$ &05 40 08.51&$-$02 45 50.6&              &              &21.85$\pm$0.22&1.91$\pm$0.44&22.80&low       \\
S\,Ori\,J053932.4$-$025220$^{\mathrm{c}}$ &05 39 32.43&$-$02 52 20.2&21.99$\pm$0.22 &1.85$\pm$0.40&21.91$\pm$0.22&2.13$\pm$0.40&22.87&high    \\
S\,Ori\,J053732.3$-$022842 &05 37 32.32&$-$02 28 42.1&              &              &22.09$\pm$0.23&1.98$\pm$0.37&23.09&low       \\
S\,Ori\,J053816.3$-$022727 &05 38 16.38&$-$02 27 27.1&              &              &22.18$\pm$0.24&1.85$\pm$0.37&23.20&low       \\
\hline
\end{tabular}
\begin{list}{}{}
\item[$^{\mathrm{a}}$] Estimated from the $z'$ photometry.
\item[$^{\mathrm{b}}$] Also in  B\'ejar et al$.$ (\cite{bejar06}).
\item[$^{\mathrm{c}}$] Also in  Caballero et al$.$ (\cite{caballero06}).
\end{list}
\end{table*}

\begin{figure*}
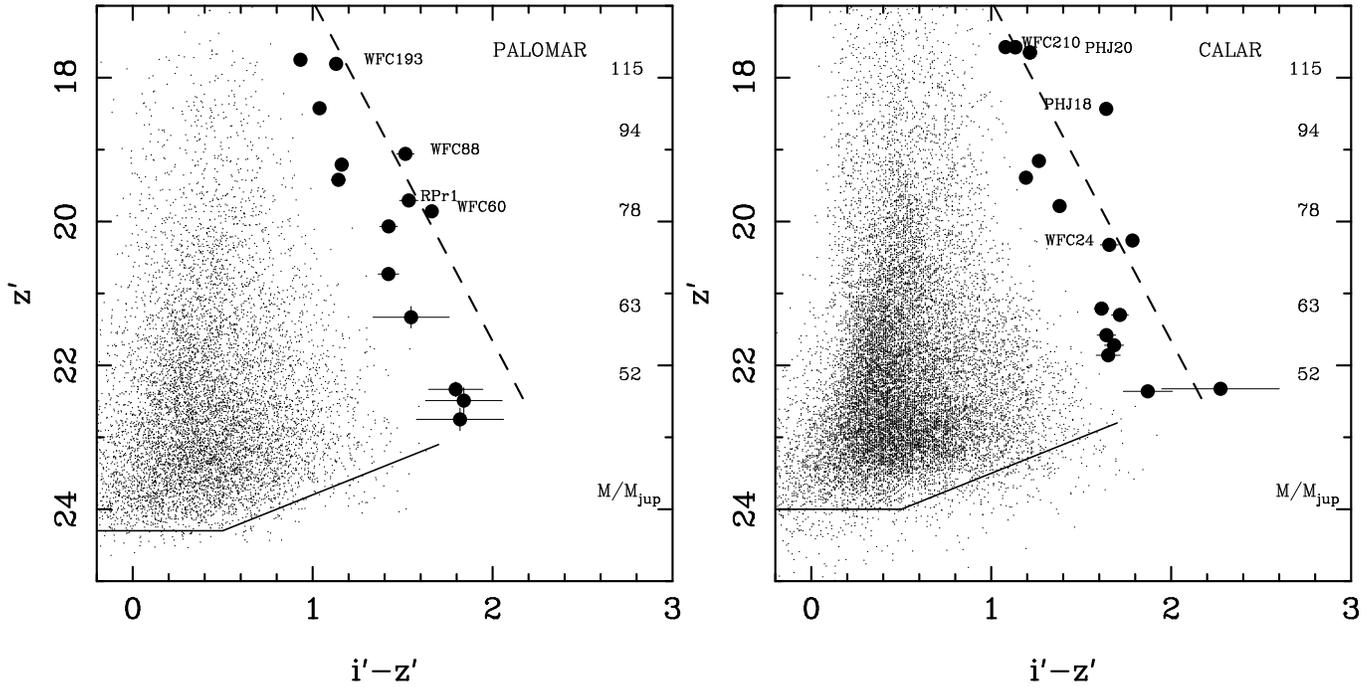
 
\resizebox{\hsize}{!}{\includegraphics{LFCPR_2paper_last.ps} \includegraphics{LAICAPR_2paper_last.ps}}
\caption{Optical color-magnitude diagramas of our Palomar (left) and Calar (right) surveys in the Praesepe star cluster. Selected Praesepe candidates are plotted as filled circles. Previously known members, which are labeled, define the cluster photometric sequence depicted with a dashed line. The solid lines indicate the limiting magnitudes of our surveys; completeness is about 1 mag brighter. Masses for an age of 500 Myr are indicated to the right side of the diagrams. Note that only PSF photometric errors are plotted. }  
\label{prae1}
\end{figure*}
 
\begin{figure*}  
\resizebox{\hsize}{!}{\includegraphics{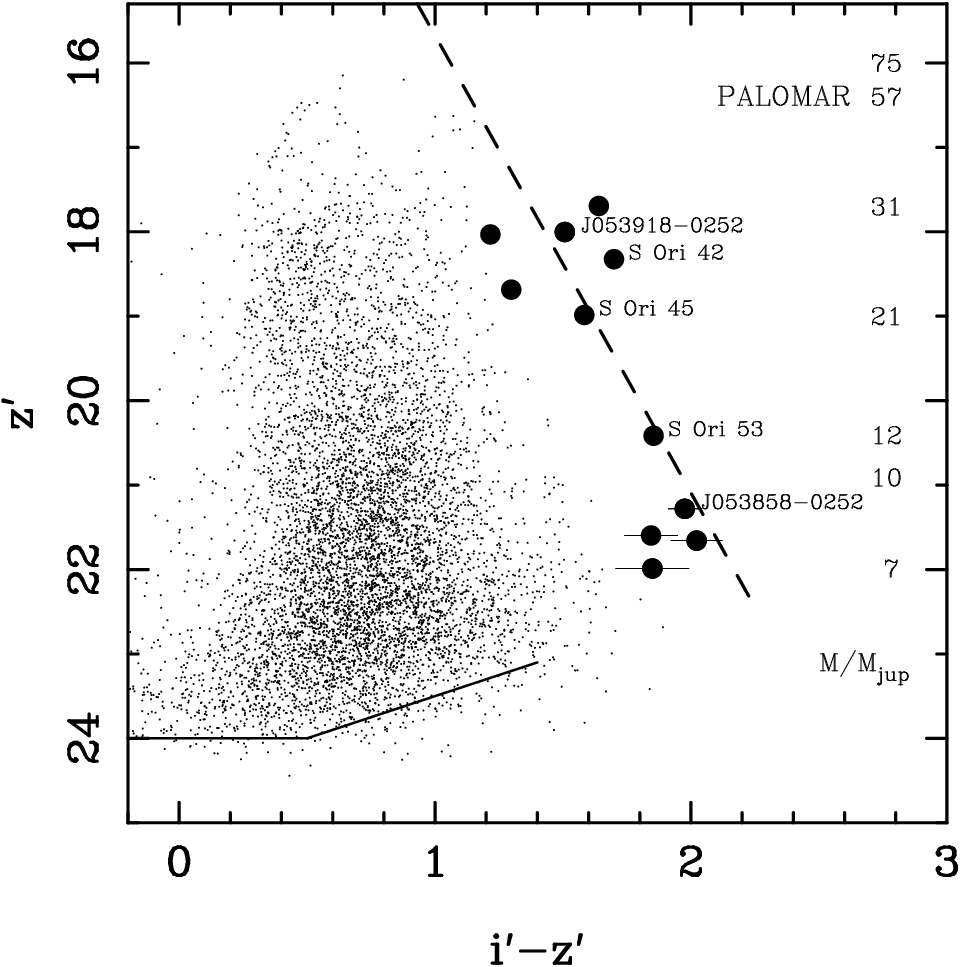}\includegraphics{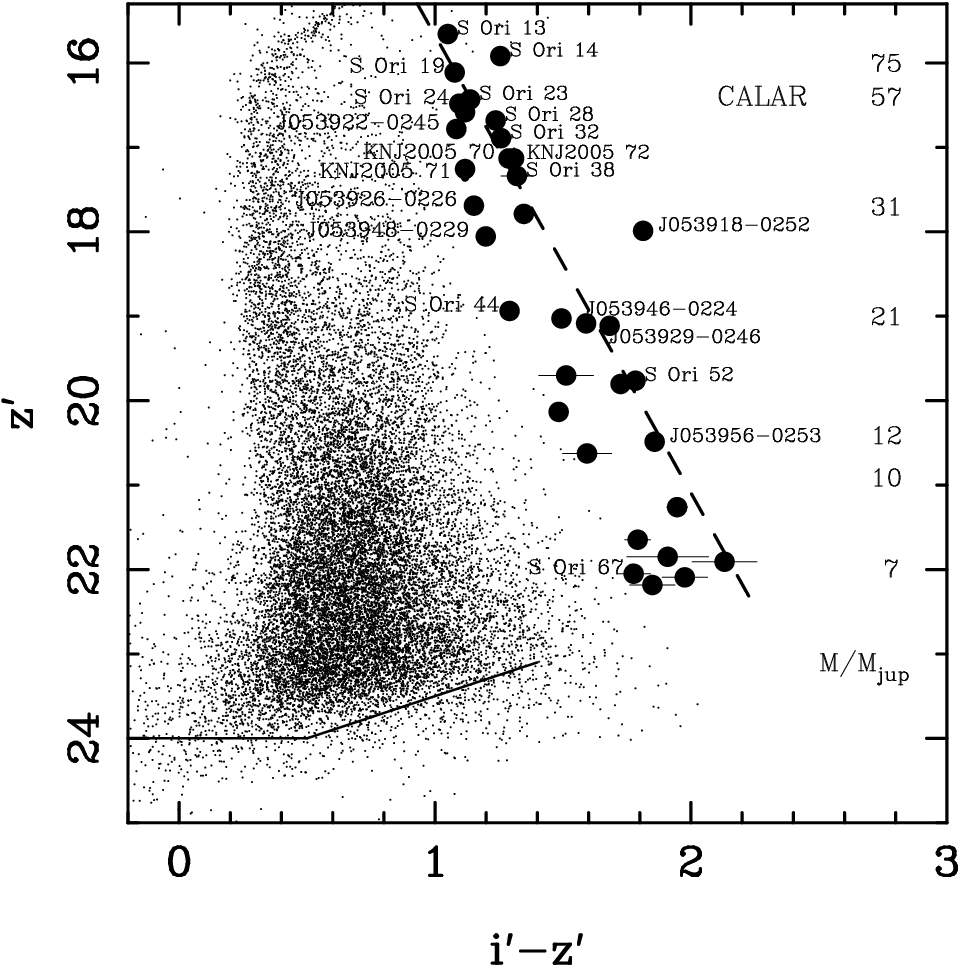}}
\caption{Optical color-magnitude diagramas of our Palomar (left)  and Calar (right) surveys in the \so~star cluster. Selected \so~candidates are plotted as filled circles. Previously known members, which are labeled, define the cluster photometric sequence depicted with a dashed line. The solid lines indicate the limiting magnitudes of our surveys; completeness is about 1 mag brighter. Masses for an age of 3 Myr are indicated to the right side of the diagrams. Note that only PSF photometric errors are plotted.}  
\label{so1}
\end{figure*}

The photometric analysis of both the Palomar and Calar surveys was performed using routines within the DAOPHOT package of IRAF\footnote{IRAF is distributed by National Optical Astronomy Observatories,  which is operated by the Association of Universities for Research in  Astronomy, Inc., under contract to the National Science Foundation, USA}. We have obtained aperture and PSF fitting photometry for all sources using circular apertures at least four times larger than the average seeing of the frames. Instrumental magnitudes have to be converted into observed magnitudes. Unfortunately, we did not observe any flux standard star in the $i'$ and $z'$ filters in any of the observing campaings. As on January 2006, no public data overlapping the regions of our survey were available to us from the Sloan Digital Sky Survey. Hence, we calibrated instrumental photometric data using previously identified Praesepe and \so~members that also appear in the Palomar and Calar surveys. For these objects, Magazz\`u et al. (\cite{magazzu98}), Hodgkin et al. (\cite{hodgkin99}), Zapatero Osorio et al. (\cite{osorio00}), B\'ejar et al. (\cite{bejar04}), Chappelle et al. (\cite{chappelle05}), Kenyon et al. (\cite{kenyon05}), and Caballero et al. (\cite{caballero04}, \cite{caballero06}) provide $I$-band photometry, and for many the $(R-I)$ color is also available. Tables~\ref{knownprae} and~\ref{knownso} show the complete list of known cluster members in our survey along with their $I$-band magnitudes taken from the literature and from the very recent calibration of B\'ejar et al. (\cite{bejar06}). We note that B\'ejar et al. (\cite{bejar06}) provide a consistent and homogeneous optical photometric calibration for a large area of the \so~cluster. 

The known objects of Tables~\ref{knownprae} and~\ref{knownso} are all red with likely spectral types cooler than mid-M. Their $I$-band magnitudes were transformed into the Sloan $i'$ and $z'$ data by applying the following equations: 
\begin{equation}
i' = I + 0.398 + 0.143 (R-I) \label{eq1}
\end{equation}
\begin{equation}
i'-z' = 1.22 (R-I) - 1.47    \label{eq2}
\end{equation}
which are valid for red objects with $(R-I) \ge$~0.75 mag (Eq.~\ref{eq1}) and $(R-I) \ge$~1.65 mag (Eq.~\ref{eq2}). Smith et al. (\cite{smith02}) provide the Sloan photometry of a large sample of equatorial standard stars drawn from the Landolt (\cite{landolt92}) catalog. We have related their $i'$ data to the Landolt $R$ and $I$ magnitudes to obtain eq.~\ref{eq1}. Equation~\ref{eq2} was derived from the relations of Smith et al. (\cite{smith02}) and the theoretical calculations by Fukugita et al. (\cite{fukugita96}). The $i'$ and $z'$ values computed for the known cluster members were then compared to the objects' instrumental magnitudes from our search, and used to define the zero point calibration of the data. We note that because of this procedure, which strongly relies on the colors of the calibrating sources, the $i'$ and $z'$ magnitudes of objects with blue colors are not reliable at all in our work. However, our main interest is the study of the very red population of the Praesepe and \so~clusters, for which the calibration should be reasonable. The \so~Palomar survey was calibrated using objects from the overlapping region with the Calar survey. The $i'$- and $z'$-band dispersion of the adopted photometric calibration for the two clusters turns out to be $\pm$0.2 mag, and the error associated to the zero points is 0.06 mag. The consistency of the calibration between the Palomar and Calar surveys in Praesepe was checked by examining the sources in the overlapping area: for any of the filters, we found no obvious offset larger than the uncertainty of the calibration. The relatively large dispersion in the obtention of the zero points is probably due to the use of various photometric sources for the $I$-band magnitudes, to the dispersion introduced by the transformation eqs.~\ref{eq1} and~\ref{eq2}, and to the possible intrinsic variability of some objects. 

We provide in Tables~\ref{knownprae} (Praesepe) and~\ref{knownso} (\so) the final $i'$ and $z'$ magnitudes of the cluster members in common with the literature as obtained from our survey. Photometric error bars account for both the PSF error provided by the IRAF routines and the uncertainty in the photometric calibration. Limiting (4--5-$\sigma$) and completeness (10-$\sigma$) magnitudes for each filter and survey are provided in Table~\ref{optlog}. In short, our search may be sensitive to objects down to $i'$\,=\,24.5 and $z'$\,=\,24 mag, while the survey completeness is determined at $i'$\,$\sim$\,23.5 and $z'$\,$\sim$\,23 mag. However, because we impose that the cluster candidates are detected in both filters and the faintest cluster members should have very red colors ($i'-z' \sim 2$), our survey is actually limited by the blue $i'$-band.

\subsection{Identification of optical candidates}

The color-magnitude diagrams of our optical survey are illustrated in Figs.~\ref{prae1} (Praesepe) and~\ref{so1} (\so). We identify the previously known cluster members with labels in the Figures. These objects are clearly separated from the field sources in the diagrams, and define a photometric sequence characteristic of each cluster. The extrapolation of the $z'$ vs$.$ $(i'-z')$ cluster sequence down to the limit of our survey is plotted as dashed lines. As can be seen from the Figures, the slope of the Praesepe and \so~dashed lines is quite similar provided the fact that true cluster members become increasingly redder for fainter magnitudes. The limiting magnitudes of both filters are also indicated by solid lines in the Figures. 

To select cluster member candidates we have plotted an imaginery straight line that is parallel to the cluster photometric sequence and shifted 0.4 mag towards the blue. For bluer colors the number of detections increases very drastically, probably indicating that these are field objects. Additionally, this line coincides with the apparent lower envelope delineated by the previously known candidates of the \so~cluster (e.g., see the location of S\,Ori\,44 and S\,Ori\,67 in Fig.~\ref{so1}). The photometric dispersion observed in other optical colors among low-mass proper motion members of the Pleiades and Praesepe clusters is about 0.6--0.7 mag. 

\begin{table*}
\caption{Near-infrared photometry of a few new Praesepe candidates.}
\label{prae_nir_phot}           
\centering           
\begin{tabular}{llcccccc}
\hline    
\hline    
Object                                  & Teles. & $J$            & $H$            & $K_s$          & $z'-J$        & $J-K_s$       & Phot. prob.$^{\mathrm{a}}$ \\
\hline    
Prae\,J083850.6$+$192317$^{\mathrm{b}}$ & 2MASS  & 16.37$\pm$0.11 & 16.01$\pm$0.18 & 15.40$\pm$0.19 & 1.20$\pm$0.23 & 0.97$\pm$0.12 & high / high\\
Prae\,J084051.2$+$191010$^{\mathrm{c}}$ & 2MASS  & 16.75$\pm$0.15 & 15.76$\pm$0.15 & 15.59$\pm$0.23 & 1.00$\pm$0.25 & 1.16$\pm$0.27 & low  / high\\
WFC\,24$^{\mathrm{d}}$                  & TCS    & 17.71$\pm$0.33 &                &                & 2.6$\pm$0.4   &               & high / high\\
Prae\,J084111.9$+$192043                & WHT    & 19.85$\pm$0.07 & 18.95$\pm$0.11 & 18.28$\pm$0.10 & 2.48$\pm$0.21 & 1.57$\pm$0.12 & high / high\\
\hline    
\end{tabular}
\begin{list}{}{}
\item[$^{\mathrm{a}}$] Optical / near-infrared photometric cluster membership probability.
\item[$^{\mathrm{b}}$] 2MASS\,J08385059$+$1923164.
\item[$^{\mathrm{c}}$] 2MASS\,J08405122$+$1910106.
\item[$^{\mathrm{d}}$] $J$\,=\,17.99$\pm$0.03, $H$\,=\,17.30$\pm$0.02, $K$\,=\,16.78$\pm$0.02 (Chappelle et al. \cite{chappelle05}).
\end{list}
\end{table*}

\begin{table*}
\caption{Near-infrared photometry of new \so~candidates.}
\label{sigori_nir_phot}           
\centering           
\begin{tabular}{lcccl}
\hline    
\hline    
Object                                    & $J$            & $z'-J$        & Phot. prob.$^{\mathrm{a}}$ & Ref. \\
\hline    
S\,Ori\,J053724.5$-$021856                & 15.36$\pm$0.09 & 1.22$\pm$0.23 & high / high                & B\'ejar et al. (\cite{bejar06})\\
S\,Ori\,J053734.5$-$025559$^{\mathrm{b}}$ & 15.71$\pm$0.06 & 2.08$\pm$0.22 & high / high                & 2MASS \\
S\,Ori\,J053855.0$-$024034$^{\mathrm{c}}$ & 15.92$\pm$0.07 & 1.77$\pm$0.22 & high / high                & 2MASS \\
S\,Ori\,J053908.4$-$023902$^{\mathrm{d}}$ & 16.46$\pm$0.10 & 1.57$\pm$0.23 & low  / high                & 2MASS \\
S\,Ori\,J053731.9$-$021634                & 17.46$\pm$0.08 & 1.57$\pm$0.23 & high / low                 & B\'ejar et al. (\cite{bejar06})\\
S\,Ori\,J053913.9$-$021621                & 17.57$\pm$0.10 & 2.23$\pm$0.24 & high / high                & B\'ejar et al. (\cite{bejar06})\\
S\,Ori\,J053813.5$-$021934                & 18.56$\pm$0.20 & 1.57$\pm$0.28 & low  / low                 & B\'ejar et al. (\cite{bejar06})\\
S\,Ori\,J053844.5$-$025514                & 19.31$\pm$0.08 & 2.35$\pm$0.28 & high / high                & Caballero (\cite{tesis06})\\
S\,Ori\,J054008.5$-$024551                & 19.43$\pm$0.16 & 2.43$\pm$0.27 & low  / high                & Caballero et al. (\cite{caballero06})\\
S\,Ori\,J053932.4$-$025220                & 19.54$\pm$0.08 & 2.37$\pm$0.24 & high / high                & Caballero (\cite{tesis06})\\
S\,Ori\,J053919.2$-$025910                & 19.83$\pm$0.09 & 1.77$\pm$0.29 & low  / low                 & Caballero et al. (\cite{caballero06})\\
\hline    
\end{tabular}
\begin{list}{}{}
\item[$^{\mathrm{a}}$] Optical / near-infrared photometric cluster membership probability.
\item[$^{\mathrm{b}}$] 2MASS\,J05373457$-$0255589, $H$ = 14.86$\pm$0.06, $K_s$ = 14.39$\pm$0.08 mag.
\item[$^{\mathrm{c}}$] 2MASS\,J05385492$-$0240338, $H$ = 15.17$\pm$0.06, $K_s$ = 14.71$\pm$0.11 mag.
\item[$^{\mathrm{d}}$] 2MASS\,J05390846$-$0239013, $H$ = 15.72$\pm$0.10, $K_s$ = 15.19$\pm$0.13 mag.
\end{list}
\end{table*}

A total of 28 candidates have been extracted from our survey of the Praesepe cluster, 20 of which are new. Regarding \so, 43 candidates have been found to match our selection criteria, 18 of which have no previous mention in the literature. Tables~\ref{newprae} and~\ref{newso} provide IAU names, coordinates and the optical photometry of all the new candidates. The astrometric plate solution has been obtained for all the CCDs by comparison of the observations with the USNO B-1 catalog (Monet et al. \cite{monet03}). Possible spatial distortions have been modelled in a standard astrometric reduction procedure using the task PLTSOL of IRAF. The quadratic mean ($rms$) of the deviations between the catalog and the computed coordinates for the transformation stars is $\pm$0.6$''$ (right ascension) and $\pm$0.5$''$ (declination) for both the Palomar and Calar surveys.


\begin{figure*}
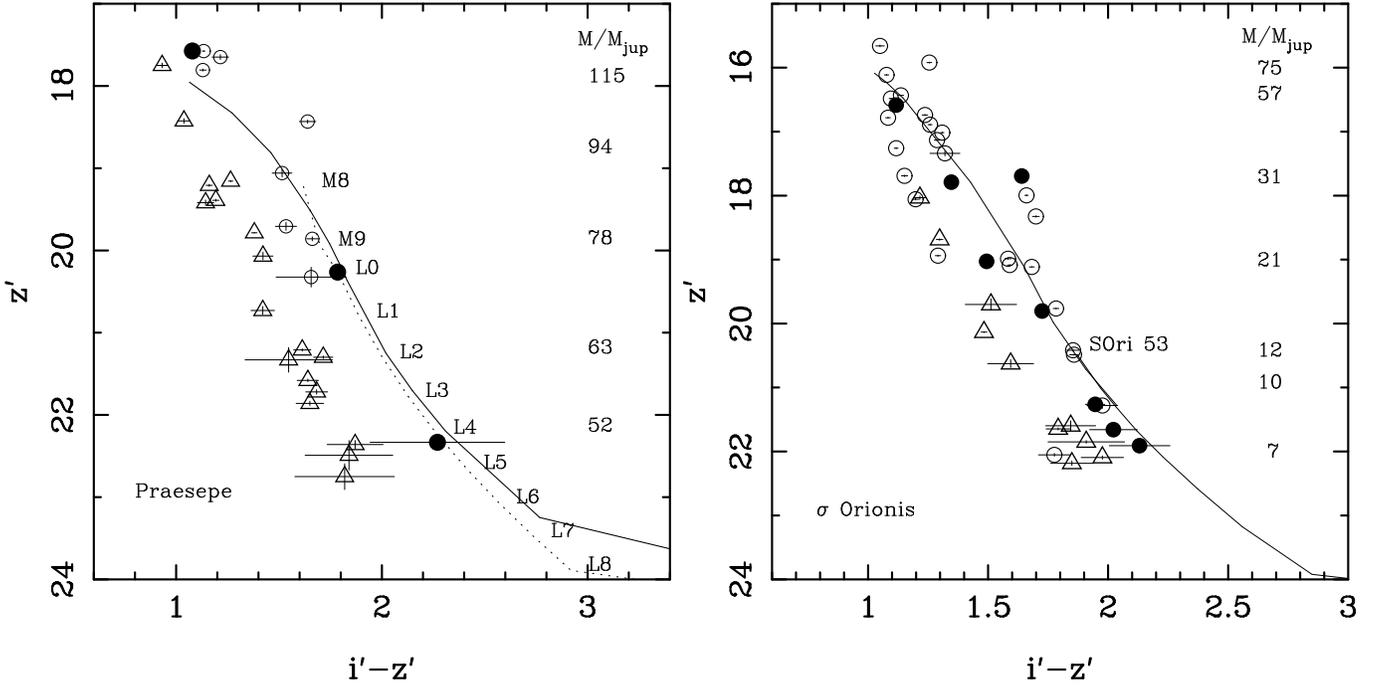
 
\resizebox{\hsize}{!}{\includegraphics{iz_prae.ps} ~ \includegraphics{iz_so.ps}}
\caption{Our Praesepe (left panel) and \so~(right panel) candidates are plotted in the optical color-magnitude diagram. Filled circles stand for the newly discovered cluster candidates with high membership probability, while previously known cluster members are plotted as open circles. Our ``less'' probable member candidates are depicted with open triangles (they lie in a sequence that appears slightly bluer than those of the clusters). Overplotted onto the data are the 500-Myr (Praesepe) and 3-Myr (\so) isochrones (solid lines) from the Lyon group (Chabrier et al. \cite{chabrier00}; Baraffe et al. \cite{baraffe98}, \cite{baraffe03}). Masses in Jovian units are indicated to the right side of the diagrams. The field sequence of late-M and L-type dwarfs at the distance of the Praesepe cluster is plotted as a dotted line in the left panel (spectral types are labeled). Note that only PSF photometric errors are shown.}
\label{iz_cands} 
\end{figure*}

Additionally, because our selection criterion may be quite generous, for each new candidate we have asigned a flag on the cluster membership probability column of Tables~\ref{newprae} and~\ref{newso} based solely on the object's proximity to the expected optical photometric sequence. If the $(i'-z')$ color of the candidate lies within 0.2 mag from the expected index given by the dashed lines of the color-magnitude planes of Figs.~\ref{prae1} and~\ref{so1}, then the flag turns to ``high'' (indicating high photometric cluster membership probability), otherwise it is ``low''. In any case, further follow-up data are required to confirm the true membership of all the candidates in their respective clusters. The eighth column of Tables~\ref{newprae} and~\ref{newso} provide the broad-band $I$ magnitudes of the new objects, which have been estimated from their Sloan $z'$ photometry. Finder charts using the $z'$ frames for both ``high'' and ``low'' probable candidates are provided in Figs.~\ref{fc_high_prae} to~\ref{fc_low_sigori}.

A few new Praesepe and \so~candidates have been found in the overlapping regions of the surveys. In general, the Calar and Palomar photometry do agree well, except for the faintest candidates (S\,Ori\,J053932.4$-$025220 and Prae\,J084111.9$+$192043), where large discrepancies (up to 0.45 mag) can be observed in the $i'$-band data; on the contrary, the $z'$ magnitudes are quite alike within less than 1-$\sigma$ the errors. This is explained by the fact that these objects appear to be notably red and have $i'$-band magnitudes indeed very close to the detection limit of our survey, where we expect large photometric uncertainties.

\subsection{Near-infrared photometry}

A few of the brightest new Praesepe and \so~candidates of Tables~\ref{newprae} and~\ref{newso} were detected by the 2 Micron All-Sky Survey (2MASS, Skrutskie et al. \cite{skrutskie06}). Their $JHK_s$ magnitudes and associated error bars are given in Tables~\ref{prae_nir_phot} and~\ref{sigori_nir_phot}. The 2MASS photometry shows rather large uncertainty because these objects are close to the detection limit of the all-sky survey. Nevertheless, both optical and near-infrared data support them as red sources.

Additionally, we carried out point observations of two of our optical Praesepe candidates using the $J$, $H$, and $K_s$ filters and the 1.5-m Carlos S\'anchez Telescope (TCS, Teide Observatory) and the 4.2-m William Herschel Telescope (WHT, Roque de los Muchachos Observatory). The date of the observations were as follows: 2005 March 07 at the TCS, and 2006 March 20 and 22 at the WHT. We used the HgCdTe, 1$''$-pixel detector at the Cassegrain focus of the TCS, and the LIRIS camera (equipped with a 0.25$''$-pixel, 1024$\times$1024 HAWAII detector) at the WHT. Observations were performed in the broad-band $J$, $H$, and $K_s$ filters. The total integration time was 2400~s for WFC\,24 ($J$-band), 640~s ($J$) and 800~s ($H$, $K_s$) for Prae\,J084111.9$+$192043, where each final image actually consisted of many (32--240) short individual exposures of 5--20~s. Raw near-infrared data were reduced in a standard way for these wavelengths, which include sky substraction and flat-fielding. Weather conditions during the campaigns were always photometric at the WHT, and some cirrus were affecting the TCS observations. The photometric analysis of the two candidates was carried out using routines within IRAF. Because of the faintness of the objects, we obtained PSF fitting photometry. Instrumental magnitudes were transformed into observed magnitudes using 2MASS sources that were present within the field of view of the detectors, i.e., the targets and the calibrators were observed simultaneously. Our final meaurements are given in Table~\ref{prae_nir_phot} along with their associated error bars, which take into account the uncertainties of the PSF photometry and the photometric calibration. We note that WFC\,24 was detected quite close to the detection limit of the TCS near-infrared image. Nevertheless, our $J$-band measurement is in quite well agreement with the data by Chappelle et al. (\cite{chappelle05}) within the expected errors. 

Our second near-infrared target in the Praesepe cluster is Prae\,J084111.9$+$192043, one of the faintest and most attractive photometric Praesepe candidates found in the optical survey. The near-infrared data do confirm it as a very red, unresolved object (average seeing of the WHT observations is 0\farcs85). This object will be discussed further in the following Section.

Near-infrared photometry of eight new \so~candidates are reported by Caballero (\cite{tesis06}), Caballero et al. (\cite{caballero06}), and B\'ejar et al. (\cite{bejar06}), where the details of data obtention and reduction can be found. We summarize in Table~\ref{sigori_nir_phot} their $J$-band magnitudes and associated uncertainties. As deduced from the $(z'-J)$ colors, all objects turn out to be red, but not all of them are red enough to fit the \so~near-infrared sequence of true cluster members. Similarly to what we did using the optical data, we have also assigned a ``low'' or ``high'' near-infrared photometric cluster membership probability flag for each one of the Praesepe and \so~candidates in Tables~\ref{prae_nir_phot} and~\ref{sigori_nir_phot}. Cluster membership based on the various optical and near-infrared color-magnitude diagrams will be discussed next.


\section{Discussion}

\subsection{Cluster membership and mass of the candidates}

Our photometric survey was designed to detect in both $i'$ and $z'$ filters Praesepe brown dwarfs down to about 50 times the mass of the planet Jupiter (1~$M_{\odot}$\,=\,1047~$M_{\rm Jup}$), or 55~$M_{\rm Jup}$ (completeness), i.e., well below the substellar mass limit defined at 75~$M_{\rm Jup}$ (Chabrier \& Baraffe \cite{chabrier00}). This prediction is based on the state-of-the-art evolutionary models by Baraffe et al. (\cite{baraffe98}), an age of 500~Myr, and a distance modulus of $(m-M)$\,=\,6.28 mag (180~pc, Robichon et al. \cite{robichon99}). For the older age of 1~Gyr, the mass limit of our survey is estimated at 60~$M_{\rm Jup}$, still in the substellar regime. In terms of spectral type, these brown dwarfs would show mid- to late-L spectral appearance with surface temperatures of about 1500\,K (Vrba et al. \cite{vrba04}). Similarly, our \so~survey is sensitive to the same spectral interval, which corresponds to masses of 6--7~$M_{\rm Jup}$ (limit) and 8~$M_{\rm Jup}$ (completeness) at the age (3\,Myr) and distance (352~pc) of the cluster. 

\begin{figure} 
\resizebox{\hsize}{!}{\includegraphics{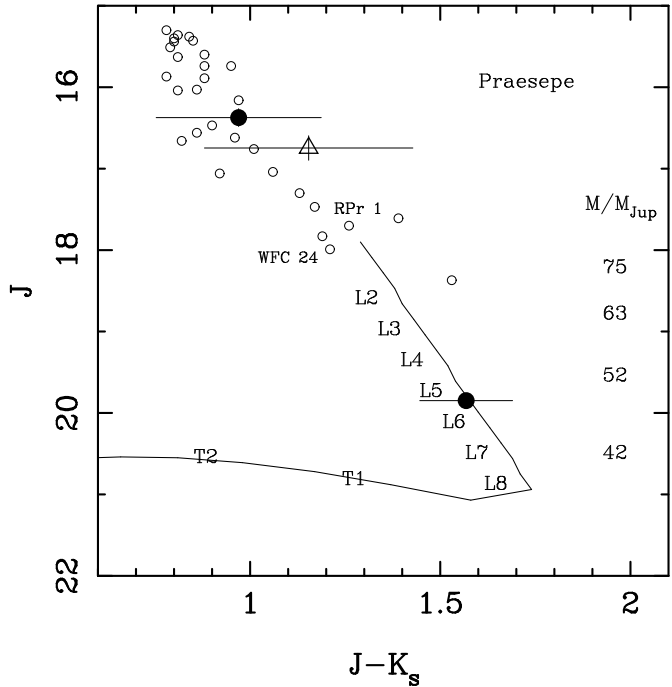}}
\caption{Near-infrared color-magnitude diagram of Praesepe members. Our ``low'' and ``high'' new optical candidates with $JK_s$ photometry are plotted as open triangles and filled circles, respectively, along with their associated error bars. Previously known members and data from the literature are shown with open circles. The field sequence at the distance of the cluster is plotted as a solid line (spectral types are labeled). Masses according to an age of 500~Myr and the models by Chabrier et al. (\cite{chabrier00}) are indicated to the right side of the diagram.}
\label{prae3} 
\end{figure}

\begin{figure} 
\resizebox{\hsize}{!}{\includegraphics{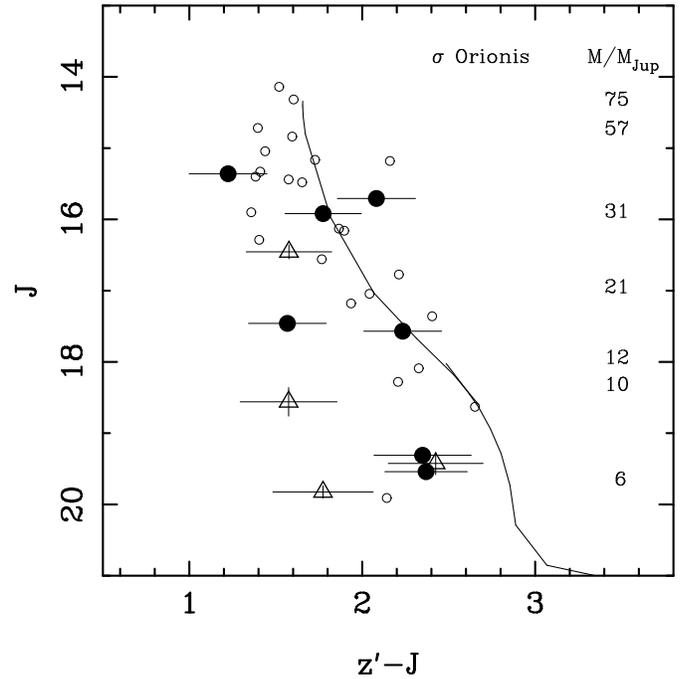}}
\caption{Optical and near-infrared color-magnitude diagram of \so~members. Our candidates with $i'z'$ high photometric membership probability are plotted as filled circles, while open triangles stand for low probability ones. Previously identified cluster members are shown with open circles. Overplotted onto the data is the 3-Myr isochrone from Chabrier et al. (\cite{chabrier00}) and Baraffe et al. (\cite{baraffe03}). Masses are labeled to the right.}
\label{sigori3} 
\end{figure}

\begin{figure*} 
\caption{Images ($2' \times 2'$ in extent) of Prae\,J084111.9$+$192043 taken at four different wavelengths. [Provided in jpeg format in astro-ph.]}
\label{J084111.9} 
\end{figure*}

There are various new Praesepe and \so~candidates discovered in our survey that have been labeled with a ``high'' flag in the optical photometric cluster membership columns of Tables~\ref{newprae} and~\ref{newso}. All of them are plotted as filled circles in Fig.~\ref{iz_cands}, where previously known cluster members from the literature are also shown. Overplotted onto the data are the 500-Myr (Praesepe) and 3-Myr (\so) isochrones obtained from the NextGen (3000--2000~K, Baraffe et al. \cite{baraffe98}), dusty (2000--1300~K, Chabrier et al. \cite{chabrier00}) and COND (1300--800~K, Baraffe et al. \cite{baraffe03}) evolutionary models. To derive the predicted $i'$ and $z'$ magnitudes from the theoretical $T_{\rm eff}$'s and luminosities, we have used the relations between the observed Sloan $(i'-z')$ colors of field late-M and L dwarfs and their measured surface temperatures and $J$-band bolometric corrections compiled from the works by Knapp et al. (\cite{knapp04}), Vrba et al. (\cite{vrba04}), and Dahn et al. (\cite{dahn02}). As can be seen from Fig.~\ref{iz_cands}, the models nicely reproduce the shape and trend of the Praesepe and \so~optical photometric sequences. Given the age of Praesepe (0.5--1~Gyr), which is close to that of the field, the cluster sequence also matches within the expected photometric errors the sequence delineated by field very low-mass stars and brown dwarfs as it is shown in the Figure. The theoretical 500-Myr isochrone appears slightly overluminous as expected for a younger age. The largest differences in luminosity are observed at colors $(i'-z')$\,$\ge$\,3~mag.

Figures~\ref{prae3} and~\ref{sigori3} illustrate the near-infrared color-magnitude diagram of the Praesepe cluster and the optical-near infrared diagram of \so, respectively. As before, the field sequence moved to the distance of Praesepe is overplotted in Fig.~\ref{prae3}. It provides a good guideline of the expected photometric trend that true Praesepe members should follow at $J$ and $K_s$ magnitudes. One of the new brightest candidates, Prae\,J083850.6$+$192317, nicely fits the cluster photometric sequence at both optical and near-infrared wavelengths (see Figs.~\ref{iz_cands} and~\ref{prae3}), as also does Prae\,J084051.2$+$191010, although this candidate was labeled with a ``low'' flag in Table~\ref{newprae}. These objects are likely Praesepe stars with masses estimated at 0.10--0.15~$M_\odot$. 

Another probable photometric Praesepe candidate is Prae\,J084039.3$+$192840, which has similar brightness and colors than those of the faintest cluster members identified by Chappelle et al. (\cite{chappelle05}). Because of its similarity to RPr\,1 (Magazz\`u et al. \cite{magazzu98}) and based on its optical color, we estimate its spectral type to be between M8 and L1, i.e., in the M--L transition. According to the evolutionary models, Prae\,J084039.3$+$192840 sits on top of the expected star--brown dwarf frontier of the Praesepe cluster. 

Very interestingly, we have identified an outstanding Praesepe candidate member that is signficantly fainter and redder than any other candidate: Prae\,J084111.9$+$192043, which is detected as a very red source in both the Palomar and Calar optical surveys. As can be seen from Fig.~\ref{iz_cands}, its optical colors are consistent with mid-L spectral type, and models predict a mass of 50--60~$M_{\rm Jup}$ for it (0.5--1~Gyr). The near-infrared photometry of Table~\ref{prae_nir_phot} and the compilation of images depicted in Fig.\ref{J084111.9} confirm its cool nature. Prae\,J084111.9$+$192043 does follow the expected Praesepe cluster sequence both at optical and near-infrared wavelengths (Figs.~\ref{iz_cands} and~\ref{prae3}). We can estimate the number of L4--L6 field dwarfs that may appear in our survey with similar magnitudes to those of Prae\,J084111.9$+$192043 by taking into account the local number density of these objects in the solar neighbourhood. Cruz et al. (\cite{cruz03}) found an L0--L4 dwarf space density of 1.9$\times$10$^{-3}$ objects\,pc$^{-3}$, while for T dwarfs (which are cooler than the L-type objects), Burgasser (\cite{burgasser01}) derived a bias-corrected value of 21$\times$10$^{-3}$ objects\,pc$^{-3}$. In qualitative terms, there are more T dwarfs than L dwarfs. We will adopt a mean density of 10$\times$10$^{-3}$ dwarfs with mid-L types per cubic parsec, which is intermediate between the early-L and T classes. For the area explored in the Praesepe survey, we would have expected 0.4 such field objects contaminating our optical search, i.e., less than 1. There is a $\sim$70\%~probability that Prae\,J084111.9$+$192043 is a true Praesepe member. We also note that there is a negligible probability for this object being a background very red galaxy. On the one hand, it remains unresolved in our good-seeing WHT near-infrared images. On the other hand, Fig.~10 of Metcalfe et al. (\cite{metcalfe06}), which displays the optical near-infrared color-magnitude diagram of galaxies in the interval $H$\,=\,16--22.5 mag, shows an empty area around the location defined by the colors and magnitudes of Prae\,J084111.9$+$192043. Nevertheless, further data are needed to unambiguously confirm its cluster membership, for example, lithium detection in its optical spectrum and proper motion. 

Regarding the \so~cluster, we note that with a few exceptions, there is a one to one relation in the optical and near-infrared photometric cluster membership probability flags given in Table~\ref{sigori_nir_phot}. By considering the optical and near-infrared photometric data of both known and new \so~candidates (36 objects out of a total of 43 candidates have photometry at various wavelengths), we estimate that the success rate on the identification of likely cluster members in our survey is about 93\%~among the ``high'' candidates, and about 38\%~among the ``low'' ones. Hence, to build reliable cluster luminosity and mass functions we discard the ``low'' optical candidates unless they are proved to follow the near-infrared sequence of \so.

Of special interest is the population number in the planetary-mass regime of the \so~cluster. In our survey, we have recovered two previously known cluster candidates whose masses could be below the deuterium burning-mass limit ($z' \ge 20.5$~mag, $J \ge 18$~mag). Additionally, we have identified three high-probable and six low-probable new planetary-mass candidates. Near-infrared photometry is available for four of them (two ``high'' and two ``low''). As can be seen from Fig.~\ref{sigori3}, one low-probable planetary-mass candidate (S\,Ori\,J053919.2$-$025910) does not show an infrared color consistent with membership in \so, and thus, it is not considered further in this paper. On the other hand, S\,Ori\,J053844.5$-$025514 and S\,Ori\,J053932.4$-$025220, which were given a ``high'' flag based on their optical data (see the right panel of Fig.~\ref{iz_cands}), also fit the near-infrared cluster sequence. This suggests that they are likely planetary-mass members of the \so~cluster. In addition, S\,Ori\,J054008.5$-$024551 shows infrared colors expected for true cluster members despite the fact it was originally classified as a low-probable optical candidate. 

\begin{figure*}
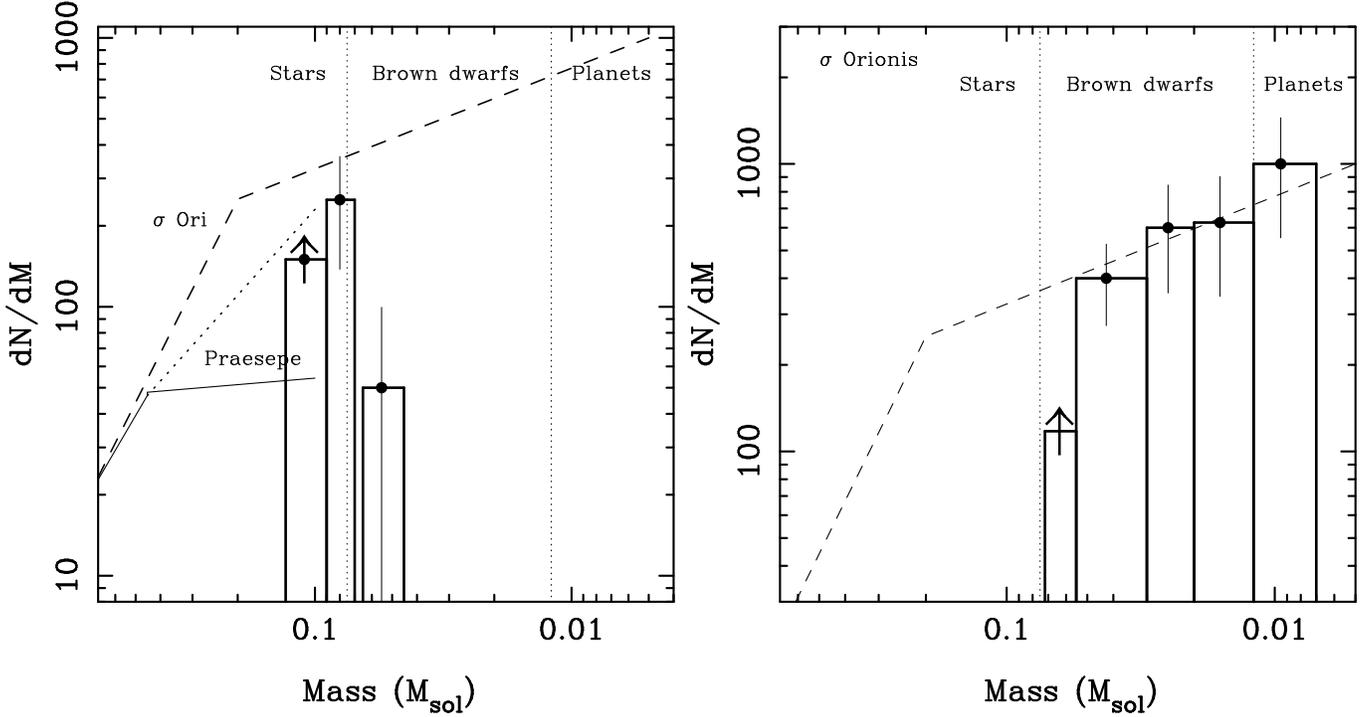
 
\resizebox{\hsize}{!}{\includegraphics{dndm_prae.ps} ~ \includegraphics{dndm_so_v2.ps}}
\caption{Praesepe (500~Myr, left panel) and \so~(3~Myr, right panel) mass functions from stars to substellar objects. The results from our surveys are depicted with filled dots and histograms. Error bars are Poissonian. In the left panel, we have also plotted the Praesepe stellar mass functions from Adams et al. (\cite{adams02}, solid line) and Hambly et al. (\cite{hambly95a}, dotted line). In both panels, and for comparison purposes, the \so~mass function of Caballero  (\cite{tesis06}) is shown with a dashed line. It is conveniently scaled to the stellar (Praesepe) and substellar (\so) regimes. Vertical dotted lines correspond to the hydrogen (0.075~$M_{\odot}$) and deuterium (0.012~$M_{\odot}$) burning-mass limits.  }
\label{mf} 
\end{figure*}

According to modern evolutionary models, S\,Ori\,J053844.5$-$025514, S\,Ori\,J053932.4$-$025220, and S\,Ori\,J054008.5$-$024551 may have masses between 6 and 10~$M_{\rm Jup}$ if their cluster membership is finally confirmed. In total, there are 6 \so~planetary candidates (6--13~$M_{\rm Jup}$) in the Palomar and Calar surveys that convey our high cluster membership probability criterion and/or fit the near-infrared cluster sequence. Another four lack follow-up near-infrared photometry. Based on the observed colors, their spectral types must range from early-L to mid-L types. From the local space number density of L1--L4 dwarfs (Cruz et al. \cite{cruz03}), we have calculated that about 2--3 field dwarf interlopers of similar spectral classifications and brightness in the interval $z'$\,=\,21.3--21.9 mag may be contaminating our \so~survey. Hence, the relatively high number of L-type cluster candidates found in our search can be understood in terms of the larger population of \so~as compared to the field.

\subsection{Praesepe and \so~substellar mass functions}

We have obtained the Praesepe and \so~mass functions by considering the recent evolutionary models of the Lyon group (Baraffe et al. \cite{baraffe03}, and references therein), cluster ages of 500 and 3~Myr, and distances of 180 and 352~pc, respectively. For each mass, the theoretical surface temperature and luminosity were converted into observables in the manner described in preceeding Sections. Only known and new candidates conveying our photometric criterion of high-probability cluster membership, and candidates that do follow the near-infrared photometric sequences of the clusters were taken into account to derive the mass functions depicted in Fig.~\ref{mf}. Thus, previously known objects like S\,Ori\,42, 44, and~67 are not included in the \so~mass function calculation. Our data are plotted as filled dots and histograms, and cover the following mass intervals: 0.045--0.13~$M_{\odot}$ (Praesepe), and 0.007--0.072~$M_{\odot}$ (\so). We note that all histogram bins of Fig.~\ref{mf} lie within the $z'$-band completeness of our search and contain at least 5 cluster member candidates with the only exception of the most massive bins, which remain unfilled due to non-linearity and saturation effects of the CCDs, and the substellar bin of Praesepe. 

As discussed by B\'ejar et al. (\cite{bejar01}), the number of \so~members appears to increase toward lower masses with a smooth transition throughout the deuterium burning-mass borderline. Overplotted in Fig.~\ref{mf} is the cluster stellar and substellar mass function recently obtained by Caballero (\cite{tesis06}). We remark the very good agreement between Caballero (\cite{tesis06}) and our derivation for the substellar regime. In the common mathematical linear representation of the mass function, d$N$/d$m$ $\sim$ $m^{-\alpha}$, our data can be fit with an exponent $\alpha$\,=\,0.6\,$^{+0.5}_{-0.1}$, where the uncertainty accounts for the different theoretical mass-luminosity relationships corresponding to an age interval 1--10~Myr. 

The substellar census of Praesepe presents a different picture. In the following discussion we assume the age of Praesepe is $\sim$500\,Myr. In the left panel of Fig.~\ref{mf}, we have included the cluster stellar mass functions from 0.1 up to 1~$M_{\odot}$ obtained by Adams et al. (\cite{adams02}) and Hambly et al. (\cite{hambly95a}). The former authors determined that the great majority of the Praesepe stars are mainly contained within a radius of 3.8~deg, while the core radius of the cluster is about 1.1~deg as derived from the King model fitting of the surface density of stars (King \cite{king62}). Our survey explores regions within the cluster core radius, where we expect the largest number of brown dwarfs to be found, and clearly extends the mass functions of Adams et al. (\cite{adams02}) and Hambly et al. (\cite{hambly95a}) toward the substellar domain. For a proper comparison between our Praesepe mass function and the results drawn from the literature, we have normalized the mass functions of Adams et al. (\cite{adams02}) and Hambly et al. (\cite{hambly95a}) by taking into account the ratio between the total cluster area and our surveyed area, and the assumption that both brown dwarfs and low-mass stars share a similar spatial distribution in Praesepe. As can be seen from Fig.~\ref{mf}, Adams et al. (\cite{adams02}) found that the number of Praesepe stars rises from 1.0 to 0.4~$M_{\odot}$, where the mass function distinctly becomes flat down to 0.1~$M_{\odot}$. This contrasts with the continuing rising function below 0.4~$M_{\odot}$ of Hambly et al. (\cite{hambly95a}). The two stellar bins derived from our search are consistent with the findings of the latter authors.

In the substellar regime, Chappelle et al. (\cite{chappelle05}) argued that there is a paucity of brown dwarfs in Praesepe leading to a turnover of the mass function near the substellar boundary. We can test this via the direct comparison of the results of our surveys in Praesepe and \so. In our search, there are 5 high-probable Praesepe candidates with masses around the substellar limit (70--90~$M_{\rm Jup}$), while only one high-probable candidate has appeared at much fainter magnitudes and with a likely mass of 52~$M_{\rm Jup}$ at 500\,Myr. On the contrary, we estimate that 4--6 Praesepe brown dwarfs should have been found in the mass interval 45--65~$M_{\rm Jup}$ if the Praesepe and \so~clusters have similar mass functions. Furthermore, the \so~photometric sequence of our survey extends well below the deuterium burning mass limit down to about 6~$M_{\rm Jup}$. The apparent drop of a factor of 5 in the number of Praesepe cluster members between the substellar limit and the brown dwarf regime is indeed indicative of the deficit of massive brown dwarfs in the central regions of Praesepe, confirming the previous conclusions of Chappelle et al. (\cite{chappelle05}). This result is also in agreement with the findings in the Hyades cluster by Dobbie et al. (\cite{dobbie02}).

It also points to a marked discrepancy between the \so~and Praesepe substellar mass functions. As can be seen from Fig.~\ref{mf}, there is a shallow rising transition between stars and brown dwarfs in \so, while the substellar bin ($\sim$55~$M_{\rm Jup}$) of Praesepe, which is affected by a strong uncertainty since it is based on one single candidate, appears significantly lower than the contiguous, more massive stellar bin. This suggests a sharp change of the Praesepe mass function at the substellar boundary to a rapidly decreasing trend for smaller masses. The population of brown dwarfs as compared to stars seems to be more abundant in \so~than in the central regions of the Praesepe cluster.

There are several explanations to account for the discrepancy between the \so~and Praesepe substellar mass functions that we will enumerate below. First, our survey covers a quite small area of the Praesepe cluster, i.e., it might yield results that are non-representative of the entire cluster population. However, we note that other surveys mapping larger areas (e.g., Chappelle et al. \cite{chappelle05}) reach similar conclusions to ours. Second, Praesepe is a much older star cluster when compared to \so. Thus, it has likely experienced some mass segregation and dynamical evolution: the smallest cluster members increase their kinetic energy and may have been dispersed from the central regions of Praesepe due to encounters and gravitational interactions with more massive cluster stars. Actually, Adams et al. (\cite{adams02}) concluded that there has been some evaporation in Praesepe, i.e., the core of the cluster is depleted of low-mass stars. Because of these effects, future searches for the Praesepe brown dwarf population should be conducted in the halo of the cluster. 

Third, it might be possible that substellar objects cool down much faster than predicted by current evolutionary models. This would make brown dwarfs fainter than expected at the age of Praesepe, i.e., there is a steeper mass-luminosity relation between the cluster smallest stars in thermonuclear equilibrium and the most massive brown dwarfs. Member candidates like Prae\,J084111.9$+$192043 would have larger masses. The substellar borderline is accurately determined to happen at M6--M7 spectral types ($\sim$2800~K) in the Pleiades (Stauffer et al. \cite{stauffer98}; Mart\'\i n et al. \cite{martin98}), which has an age of 120~Myr. It is expected that the star--brown dwarf frontier lies at cooler types for older ages. Lithium is used as a diagnostic for substellarity (Rebolo et al. \cite{rebolo92}): the detection of this atomic feature in the spectra of old, ultracool dwarfs implies lithium preservation and a mass below the substellar limit (see Basri \cite{basri00} for a comprehensive review). As discussed by Kirkpatrick et al. (\cite{kirk00}), the presence of lithium in the optical spectra of field L0-type dwarfs is $\sim$0\%, and slowly rises to 20\%~by L2, and to 50--70\%~by L6. Thus, about 50--70\%~of the field mid-L type dwarfs are bona-fide brown dwarfs for which models predict a mass smaller than about 60~$M_{\rm Jup}$. This suggests that the substellar borderline is located between spectral types M6 and L6 in both the field and the Praesepe cluster.

Other alternative possibility is the non-universality of the initial mass function. However, to address this issue a wide variety of much younger star clusters have to be studied. On the one hand, the Praesepe stellar mass function between 1 and 0.4~$M_{\odot}$ resembles that of the \so~cluster (see Fig.~\ref{mf}). On the other hand, evidence for mass segregation in the Praesepe cluster has been reported in the literature (e.g., Raboud \& Mermilliod \cite{raboud98}), which makes this cluster a non-ideal site to investigate the initial mass function.

Finally, it is possible that the age of the Praesepe cluster is significantly larger than 500\,Myr. This has an impact on the substellar mass function calculation since theoretical brown dwarf mass-luminosity relationships depend strongly on age. An older age of 800\,Myr or higher would bring the substellar mass functions of Praesepe and \so~to a better agreement. We estimate the mass of our coolest Praesepe candidate, Prae\,J084111.9$+$192043, to be 57~$M_{\rm Jup}$ at the age of 800\,Myr, and 63~$M_{\rm Jup}$ at 1\,Gyr. However, the mass bin at the substellar limit roughly changes from 500\,Myr to 1\,Gyr, suggesting a shallow transition from Praesepe very low-mass stars to massive brown dwarfs, as in \so. The various determinations of the age of Praesepe available in the literature are generally in the range 500--900\,Myr, and recent determinations based on the isochrone-fitting procedure yield a likely age of 800\,Myr (Kharchenko et al. \cite{kharchenko05}). For this and older ages, the bulk of the cluster brown dwarf population lies at magnitudes fainter than those of Prae\,J084111.9$+$192043 and will show colors typical of late-L and T-types. Future searches should take this possibility into account.



\section{Summary}

We have conducted deep photometric optical surveys in the Praesepe (0.5--1~Gyr) and \so~($\sim$3~Myr) clusters using the Sloan $i'$ and $z'$ filters, large format multi-CCD cameras, the 3.5\,m telescope on the Calar Alto Observatory and the 5-m Hale Telescope on the Palomar Observatory. Our main objective is the identification of the reddest population ($i'-z'$\,$\ge$\,1~mag) of each cluster and place constrains on the bottom end of the Praesepe and \so~mass functions. A total area of 1177 (Praesepe) and 1122 (\so) arcmin$^2$ were explored in the cluster central regions down to limiting magnitudes of $i'$\,=\,24.5 and $z'$\,=\,23.5~mag (completeness is about one magnitude brighter). Our search is sensitive to Praesepe and \so~members with masses in the intervals 0.05--0.13~$M_{\odot}$ and 0.006--0.072~$M_{\odot}$, respectively. Follow-up near-infrared $JHK_s$ photometry has allowed us to discuss the cluster membership of various candidates identified from the optical surveys. 

In \so, three new planetary-mass candidates (S\,Ori\,J053844.5$-$025514, S\,Ori\,J053932.4$-$025220, and S\,Ori\,J054008.5$-$024551) with estimated masses (0.006--0.010~$M_{\odot}$) below the deuterium burning-mass limit have been discovered. They fit the optical and near-infrared sequence of the cluster, suggesting that they are likely members of \so. From our survey, we derive a rising substellar mass function which extends from the substellar borderline to the planetary regime ($\alpha$\,=\,0.6\,$^{+0.5}_{-0.1}$, d$N$/d$M$ $\sim$ $M^{-\alpha}$), in agreement with our previous derivations (B\'ejar et al. \cite{bejar01}; Caballero \cite{tesis06}). 

In the Praesepe cluster, we have found one red candidate (Prae\,J084111.9$+$192043) whose optical and near-infrared colors coincide with our expectations for true substellar members. Its mass and spectral type are estimated at 50--60~$M_{\rm Jup}$ and L4--L6, respectively. It is the coolest and least massive Praesepe candidate reported to date. At the youngest possible ages of Praesepe (500--700~Myr), our analysis of the cluster mass function suggests a sharp change at the substellar boundary to a rapidly decreasing trend for brown dwarf masses, which clearly contrasts with the \so~mass function. This trend diminishes to a shallow transition between very low-mass stars and massive brown dwarfs when using steeper mass-luminosity relationships like those given by the models of 800\,Myr and higher.

\begin{acknowledgements}
We thank R. Rebolo for useful comments regarding this work. Partial financial support was provided by the Spanish project AYA2003-05355. This research has made use of the SIMBAD database, operated at CDS, Strasbourg, France, and has also made use of data products from the Two Micron All Sky Survey, which is a joint project of the University of Massachusetts and the Infrared Processing and Analysis Center/California Institute of Technology, funded by the National Aeronautics and Space Administration and the National Science Foundation. Based on observations collected at the Centro Astron\'omico Hispano Alem\'an (CAHA) at Calar Alto, operated jointly by the Max-Planck Institut f\"ur Astronomie and the Instituto de Astrof\'\i sica de Andaluc\'\i a (CSIC); also on observations made at the Palomar Observatory, owned and operated by the California Institute of Technology, on data collected with the William Herschel Telescope, operated on the island of La Palma by the Isaac Newton Group in the Spanish Observatorio del Roque de los Muchachos, and with the Telescopio Carlos S\'anchez operated on the island of Tenerife by the Instituto de Astrof\'\i sica de Canarias (IAC) in the Spanish Observatorio del Teide.
\end{acknowledgements}

\begin{figure*} 
\centering
\caption{Finder charts ($z'$-band, $2' \times 2'$ in extent) of Praesepe candidates labeled with a ``high'' photometric cluster membership probability.  [Provided in jpeg format in astro-ph.]}
\label{fc_high_prae}
\end{figure*}

\begin{figure*} 
\centering
\caption{Finder charts ($z'$-band, $2' \times 2'$ in extent) of Praesepe candidates labeled with a ``low'' photometric cluster membership probability.  [Provided in jpeg format in astro-ph.]}
\label{fc_low_prae}
\end{figure*}

\begin{figure*} 
\centering
\caption{Finder charts ($z'$-band, $2' \times 2'$ in extent) of \so~candidates labeled with a ``high'' photometric cluster membership probability.  [Provided in jpeg format in astro-ph.]}
\label{fc_high_sigori}
\end{figure*}

\begin{figure*} 
\centering
\caption{Finder charts ($z'$-band, $2' \times 2'$ in extent) of \so~ candidates labeled with a ``low'' photometric cluster membership probability.  [Provided in jpeg format in astro-ph.]}
\label{fc_low_sigori}
\end{figure*}

\end{document}